# Gyroelectric cubic-quintic dissipative solitons

Allan D. Boardman, Larry Velasco
Submitted September 19, 2005
Address: University of Salford, Institute of the Material research, School Comp, Science and Eng., M5 4WT, Salford, Greater Manchester, UK.
Email: a.d.boardman@salford.ac.uk

*Abstract*
The influence of an externally applied magnetic field upon classic cubic quintic dissipative solitons is investigated using both exact simulations and a Lagrangian technique. The basic approach is to use a spatially inhomogeneous magnetic field and to consider two important geometries, namely the Voigt and the Faraday effects. A layered structure is selected for the Voigt case with the principal aim being to demonstrate non-reciprocal behaviour for various classes of spatial solitons that are known to exist as solutions of the complex Ginzburg-Landau cubic-quintic envelope equation under dissipative conditions. The system is viewed as dynamical and an opportunity is taken to display the behaviour patterns of the spatial solitons in terms of two-dimensional dynamical plots involving the total energy and the peak amplitude of the spatial solitons. This action this leads to limit cycle plots that beautifully reveal the behaviour of the solitons solutions at all points along the propagation axis. The closed contour that exists in the absence of a magnetic field is opened up and a limit point is exposed. The onset of chaos is revealed in a dramatic way and it is clear that detailed control by the external magnetic field can be exercised. The Lagrangian approach is adjusted to deal with dissipative systems and through the choice of particular trial functions, aspects of the dynamic behaviour of the spatial are predicted by this approach. Finally, some vortex dynamics in the Faraday configuration are investigated.

I.      Introduction

In recent years a number of papers have appeared that focus upon the gyroelectric behaviour of spatial solitons [1-6]. This kind of work contributes to the growing interest in complex media. Material properties enter Maxwell's equations through the constitutive relationship and to move beyond purely isotropic materials involves complexity that could be the addition of some form of nonlinearity, the introduction of anisotropy or a generalized gyrotropy. Materials can be rigorously classified in terms of their behaviour and, in general, the study of bianisotropy [7] is a generic approach that includes natural chirality and forced gyrotropy [8], of the kind that is produced by an external magnetic field. Magneto-optics falls into the latter category and involves the polarisation and vector character of light [9]. Given this fact, the application of an external magnetic field to a system sustaining spatial solitons holds out the promise of applications based upon non-reciprocal gyroelectric behaviour. This is a dependence upon handedness that will be classified here into forward and backward waves that have a particular propagation direction with respect to the direction of the applied magnetic field. In some ways, magneto-optic behaviour appears to be like the natural chirality that is displayed by solutions of sugar, for example. This kind of handedness, however, is reversible, which makes it very different from the non-reciprocal behaviour associated with the magneto-optically-driven gyroelectric behaviour to be investigated here.

In an optically active medium the displacement vector is $\boldsymbol{D} = \varepsilon_0 [\boldsymbol{\varepsilon} \cdot \boldsymbol{E} + i\boldsymbol{g} \times \boldsymbol{E}]$, where $\boldsymbol{\varepsilon}$ is the relative permittivity of the medium, and $\varepsilon_0$ is the permittivity of the free space. $\boldsymbol{g} = \xi \boldsymbol{k}$ is called the *gyration vector*, the wave vector is $\boldsymbol{k}$ and $\xi$ is a constant for the purposes of this part of the discussion. Since $\boldsymbol{g}$ depends upon $\boldsymbol{k}$, in the case of optical activity, the rotation of the polarization plane, implied by the existence of term $\boldsymbol{g} \times \boldsymbol{E}$ in the constitutive relation, is reversed if the direction of $\boldsymbol{k}$, and hence the propagation direction, is reversed. In a magneto-optical case $\boldsymbol{g} = G\boldsymbol{B}_o$, where, for simplicity, $G$ is set here to be another constant, and $\boldsymbol{B}_o$ is an applied magnetic flux density. Clearly in this case $\boldsymbol{g}$ depends only upon the direction of the applied magnetic field so the rotation of the plane of polarisation around this field cannot be reversed by reversing the propagation direction. In other words, non-reciprocal behaviour can now be expected.

Magneto-optics in the linear domain is a significant global activity with a lot of emphasis on the generation of Kerr rotations [10]. There is also a considerable interest in the nonlinear Kerr rotation to be found in the second-harmonic generated by reflecting laser beams off magnetic surfaces. This investigation sets out to extend the reach of magnet-optics to include the propagation of dissipative spatial solitons, under the non-reciprocal influence of adjacent, or ambient, magneto optical materials. The latter are widely available and are in a very advanced state, so it should be possible to investigate experimentally the kind of predictions discussed below.

In the model to be developed below a wave guiding structure is used, for the Voigt effect, in which spatial solitons are created under conditions of loss and gain but with the layer sustaining the spatial solitons, being interfaced to a magneto-optic substrate composed of a classic material like the magnetic insulators called rare-earth iron garnets [10] (YIG).

The basic complex Ginzburg-Landau cubic-quintic equation is established[11,12], below, together with the relevant magneto-optical influence. This is followed up by a Lagrangian analysis designed for a dissipative medium. The numerical work, first of all, establishes nonreciprocal behaviour, using exact simulations. These are then examined in the light of the Lagrangian analysis in order to discover whether the Lagrangian work offers anything more that a qualitative or semi-quantitative guide to the exact behaviour. The paper ends with a brief account of some interesting work on the behaviour of vortices in a external magnetic field.

**II.   Basic equations**

The analysis here focuses, initially, upon an asymmetric waveguide that contains a nonlinear layer sustaining spatial solitons. The waveguide needs to be asymmetric if the Voigt effect is not to be negligible [1-6]. The nonlinear layer sits upon a magneto-optic substrate and has an air cladding. Addressing the magneto-optics first, a time-dependence $e^{-i\omega t}$, leads directly to the following equation for the electric field $\boldsymbol{E}$.

$$\nabla^2 \boldsymbol{E} + \frac{\omega^2}{c^2} \boldsymbol{\varepsilon} \cdot \boldsymbol{E} - \nabla(\nabla \cdot \boldsymbol{E}) = 0 \tag{1}$$

The relative permittivity tensor $\boldsymbol{\varepsilon}$ expresses its magneto-optic character through the existence of off-diagonal elements that are positioned within the tensor in response to the direction of the propagating waves, relative to the direction of the applied magnetic field. In this investigation, the propagation is along the z-axis and the applied magnetic field is along the z-direction for the Faraday effect, and in the (x,y) plane, but directed along the x-axis, for the Voigt effect. For these cases the relative permittivity tensors are [4]

$$\textbf{Faraday}: \varepsilon = \begin{pmatrix} n^2 & -iQn^2 & 0 \\ iQn^2 & n^2 & 0 \\ 0 & 0 & n^2 \end{pmatrix} \quad \textbf{Voigt}: \varepsilon = \begin{pmatrix} n^2 & 0 & 0 \\ 0 & n^2 & -iQn^2 \\ 0 & iQn^2 & n^2 \end{pmatrix} \tag{2}$$

in which $n$ is the magnetic field-independent refractive index of the magneto-optical material and it is assumed that all the diagonal elements are equal. This is not literally correct, for materials like YIG, for example, but it is a very good approximation [3]. The magnetisation distribution created by the applied magnetic field is given by the factor $Q$. For magnetic insulators like YIG, $Q \approx 10^{-4}$ and is often treated as a constant proportional to the magnitude of the applied magnetic field. In practice, the application of an external magnetic field can be provided by an electrode structure [1-6] that has varying degrees of complexity. This structure can be placed directly, or even grown, upon the waveguide, or placed in its near neighbourhood. Through this structure an electric current flows and the result is an inhomogeneous magnetisation distribution that makes $Q$ a function of the spatial coordinates. A simple electrode structure is a single wire carrying a current I, as shown in Fig.1.

If the wire lies along the *z*-axis then the magnetic field component that creates the magnetisation will be a vector that is tangential to circles in the $(x, y)$ plane, centred upon the wire. There are then components of the magnetisation parallel to the *x*- and *y*-axes but only the one parallel to the *x*-axis will be important because the components along y will give rise to a polar magnetic effect. The latter will cause TE-TM coupling but since the phase matching condition is not met in this type of guide, with its large birefringence the effect is entirely negligible. This investigation will adopt this method of supplying an external field and all the models introduced below will assume that $Q \equiv Q(x)$.

In the development of (1) it is important to realise that $\nabla \cdot \boldsymbol{E} \neq 0$, and that this is a pivotal issue in a forced gyrotropic medium. The full electric field in the guiding structure is defined here as $\boldsymbol{E} = (E_x, E_y, E_z) e^{(-i\omega t)}$. It is interesting that $E_z$, even though it rises rapidly with the refractive-index change across an interface in a waveguide, soon saturates and actually remains small, in comparison to the other electric field components. It is safe, therefore, to assume that for all practical purposes $E_z = 0$. For a spatial soliton beam, the *x*- and *y*-directions can be measured in units of the natural beam width $D_0$. Also, in a planar waveguide structure, only *x* will come into play, but both *x* and *y* will feature in vortex propagation as a 2D entity in the bulk. For guided waves, in the structure displayed in Fig.1, both TM and TE waves will be coupled in the Faraday configuration and phase matching might have to be addressed. It is removed here but it is not unreasonable to address this kind of "form" birefringence experimentally through some applied phase compensation, perhaps by making the applied field periodic. In the bulk this problem does not arise. The *z*-axis is scaled in diffraction, or Rayleigh, lengths that are defined as $kD_0^2$, where *k* is the bulk wave number, or the average wave number, if the $E_x$, $E_y$ components are associated with a waveguide. As can be seen from equations (1) and (2), the principal term involving *Q*, because of the scaling, attracts a factor $2k^2 D_0^2 n^2$. The final term in equation (1) introduces derivative terms, however, in the following way

$$\nabla \cdot \boldsymbol{E} = i \frac{dQ(x)}{dx} E_y + iQ(x) \frac{\partial E_y}{\partial x} - iQ(x) \frac{\partial E_x}{\partial y} \qquad (3)$$

Such terms do not scale in the same way as the leading Q term in the rest of the equations so they can be neglected. A typically modest value of Q is $Q \approx 10^{-4}$ but this can be enhanced by at least an order of magnitude by using doped YIG for example [1-6,10]. For reasonable beam widths of 8μm diameter, $Q_1 = \frac{\omega^2}{c^2} D_0^2 n^2 Q$, so the effective magneto-optic parameter is now $Q_1 \approx 1$. This is four orders of magnitude more significant than any other *Q* terms that arise as derivatives, or multipliers that are contributing to extra diffraction. $Q_1$ will now be used in the envelope equation.

For the Faraday effect, using *slowly varying* complex electric field amplitudes $\psi_{x,y}$, scaled to produce a compact dimensionless form for the final evolution equation, and then cast into a rotating coordinate system, defined as $\psi_+ = \frac{1}{\sqrt{2}} (\psi_x + i\psi_y)$, $\psi_- = \frac{1}{\sqrt{2}} (\psi_x - i\psi_y)$, a decoupled set of equations emerges. For example, when studying linear vortices in a magnetic field that is applied to a bulk nonlinear magneto-optic medium such as semi-magnetic semiconductor, or an atomic gas like sodium, or cesium, [9] the following 2D equation is appropriate

$$i\frac{\partial \psi_+}{\partial z} + \frac{1}{2}\left(\frac{\partial^2 \psi_+}{\partial x^2} + \frac{\partial^2 \psi_+}{\partial y^2}\right) + Q_1\psi_+ = 0 \tag{4}$$

and the x and y wave number components are equal. Thus, for this example, an arbitrarily polarised input to a Faraday system evolves, in the linear case, as uncoupled counter-rotating, circularly polarised waves. Equation (4) is sufficient to describe 2D vortex motion and by dropping down to 1D can be used to study phase-matched operations in a waveguide.

Still within the linear domain [3], the *Voigt configuration* requires an asymmetric waveguide to produce magneto-optic effects to $O(Q)$, because a symmetrical structure leads only to effects $O(Q^2)$, as would be the case in free space. For the guiding structure shown in Fig.1

$$\boldsymbol{E} = \psi\left[\hat{\boldsymbol{y}}\xi_y(y) + \hat{\boldsymbol{z}}\xi_z(y)\right]e^{i(\omega t - \beta z)} \tag{5}$$

where, once again, $\psi$ is a slowly varying amplitude, and $\xi_y(y)$, $\xi_z(y)$ are the linear modal field components appropriate to the waveguide. The purely magneto-optic effect, leaving out diffraction, for the moment, is modelled by

$$i\frac{\partial \psi}{\partial z} = \frac{\omega}{c}\bar{\varepsilon}_{yz}\psi \tag{6}$$

where $\varepsilon_{yz} = -Qn^2$ but here the role of magneto-optics comes in through the following average over the whole waveguide structure.

$$\bar{\varepsilon}_{yz} = \frac{c}{\omega\beta^2}\frac{\int \varepsilon_{yz}\xi_y\left(\frac{\partial \xi_y}{\partial y}\right)dy}{\int\left(|\xi_y|^2 + |\xi_z|^2\right)dy} \tag{7}$$

It can be seen from (7) that $\bar{\varepsilon}_{yz} = 0$ for a symmetric guide and also that the diffraction of the beam is, once again, easily obtained by including a $\frac{\partial^2 \psi}{\partial^2 x}$ term in the envelope evolution equation [1-6].

The foregoing has addressed two popular magneto-optic effects, in the absence of any nonlinear, or absorptive/gain, processes. These can now be included through the adoption of the cubic-quintic complex Ginzburg-Landau model[13]. The final generic equation, scaled in precisely the same way as before, now incorporates the magneto-optic effects discussed above, and is

$$i\frac{\partial \psi}{\partial z}+i\alpha\psi+\left(\frac{1}{2}-i\beta\right)\left(\frac{\partial^2 \psi}{\partial x^2}+\frac{\partial^2 \psi}{\partial y^2}\right)+(1-i\varepsilon)|\psi|^2\psi-(\nu-i\mu)|\psi|^4\psi+Q_1(x)\psi=0 \quad (8)$$

This equation only reaches this form after an arbitrary transformation of the amplitudes. However, the length scales are the same for both magneto-optic configurations, so scaling the amplitudes in a slightly different way in each case [14] can have no effect upon the solitons evolution patterns. In other words, the details of the amplitude transformations is unimportant because all the control of the solitons behaviour will be exercised through the parameters that have now been introduced.

Specifically, $\alpha$ is a measure of the linear damping, $\beta$ represents the possibility of diffusion, $\varepsilon$ measures the cubic power gain, $\mu$ is a quintic loss term and $\nu$ is a self-defocusing contribution to the beam evolution. $Q_1(x)$ is now a generalised magneto-optic parameter that is also a function of $x$. The signs of $\alpha$, $\mu$ and $\nu$ in (8) are assigned through the data, later on, to give published [15-17] starting points that will be used to assess the effect of the applied magnetic field distribution. For a study of magneto-optic vortex solitons, equation (8) will be used as it stands and is interpreted as propagation in a 2D bulk medium of circularly polarised modes within a Faraday configuration. Using the scaling adopted earlier, the magneto-optic parameter for this case is $Q_1=\frac{\omega^2}{c^2}D_0^2 n^2 Q$. The Voigt effect in a planar waveguide can be recovered from (8) by omitting the term $\frac{\partial^2 \psi}{\partial y^2}$, adopting the same scaling, and setting the magneto-optic parameter to $Q_1=\frac{\omega^2}{c^2}D_0^2 n^2 \bar{\varepsilon}_{yz}$. Note that, for the 1D case, an integration over the guide that deploys the modal structure must be made. In both the Faraday and Voigt cases, however, $Q_1 \propto Q$ and the both the scaling transformations supply a factor of $O(10^4)$. Hence the $Q_1$ factor will be of a similar order of magnitude in both cases and will always have a significant influence. The earlier elegant discussions of cubic-quintic dissipative solitons [15-18] have shown already that it is possible to generate significantly distinct classes of solutions in the absence of any magneto-optic effect, so the introduction of nonreciprocal behaviour through an applied magnetic field should dramatically alter the solitons behaviour. It is important to emphasise that $Q_1$ is the same order of magnitude in each application, even though $Q_1$ in the Voigt case involves a modal averaging factor arising from the guiding in the asymmetric planar structure. This modal factor is the order of unity and will "switch off" $Q_1(x)$ if the waveguide is symmetric.

### III. Lagrangian Analysis

The study of solitons dynamics can often be based upon variational theory. Any analytical progress depends upon the adoption of the average variational method first introduced by Whitham [19,20]. It centres upon the use of an average Lagrangian density and the inevitable introduction of trial functions. Whitham points out that the variational approach is not a separate method and that it permits the development of quite general results but naturally the accuracy of any description must depend upon a judicious choice of trial functions. For the dissipative cubic-quintic system considered here, the Lagrangian density is

$$\mathcal{L} \equiv \mathcal{L}\left(\psi, \psi^*, \frac{\partial \psi}{\partial z}, \frac{\partial \psi^*}{\partial z}, \frac{\partial \psi}{\partial x}, \frac{\partial \psi^*}{\partial x}\right) \tag{9}$$

where the complex amplitude function $\psi(x,z)$ is just a function of x and z. Whitham showed that

$$\delta L = \delta \int_{-\infty}^{\infty} \mathcal{L} dx = 0 \tag{10}$$

where $L$ is the averaged Lagrangian[19-24].

Powerful applications of this have been made to conservative systems [21-24] that are free of absorption and gain but, even if there is simple linear gain, a transformation can be made that reduces the problem once again to the familiar form used for conservative systems. It is necessary, therefore, to extend the reasoning behind the application of the Whitham average Lagrangian procedure, in order to make it applicable, in principle, to the complex Ginzburg-Landau cubic-quintic dissipative models.
In the absence of gain or absorption, the conservative gyroelectric Lagrangian density is

$$\mathcal{L} = \frac{i}{2}[\psi \frac{\partial \psi^*}{\partial z} - \psi^* \frac{\partial \psi}{\partial z}] + \frac{1}{2}\frac{\partial \psi}{\partial x}\frac{\partial \psi^*}{\partial x} - \frac{1}{2}\psi^2 \psi^{*2} + \frac{v}{3}\psi^3 \psi^{*3} + Q(x)\psi\psi^* \tag{11}$$

so the final envelope equation for the gain/loss *non-conservative* system is

$$\frac{\delta \mathcal{L}}{\delta \psi^*} = R \tag{12}$$

where

$$R = -i\alpha\psi + i\beta\frac{\partial^2 \psi}{\partial x^2} + i\varepsilon|\psi|^2\psi - i\mu|\psi|^4\psi \tag{13}$$

It is equally valid, however, to generate the envelope equation from the complex conjugate of (12). Hence in the case arising because of the presence of absorption and gain, and any other non-conservative processes, the full functional variation of the conservative Lagrangian density $\mathcal{L}$ should be used. This variation, by definition, is

$$\delta \mathcal{L} = \frac{\delta \mathcal{L}}{\delta \psi} \delta \psi + \frac{\delta \mathcal{L}}{\delta \psi^*} \delta \psi^* = R^* \delta \psi + R \delta \psi^* = 2\Re[R\delta \psi^*] \qquad (14)$$

where $\Re$ denotes the taking of the real part.

This step can now be taken further by progressing to the use of L, the average *conservative* Lagrangian, by first of all adopting a trial function $\psi_T$ that has z-dependent parameters. The latter will be properties of the solitons, such as amplitude ($f_1$), width ($f_2$), velocity($f_3$), position($f_4$), phase ($f_5$), or indeed any other feature that needs to be emphasised. For a conservative system, the average Lagrangian is used with Euler-Lagrange equations that are modified to be a variation with respect to the z-dependent parameters $f_j$, i.e.

$$\frac{\delta L}{\delta f_j} = \frac{d}{dz}\left(\frac{\partial L}{\partial\left(\frac{\partial f_i}{\partial z}\right)}\right) - \frac{\partial L}{\partial f_i} - = 0 \qquad (15)$$

Since the x-dependence is integrated out when forming the average Lagrangian only the z coordinate-dependence remains so that d/dz replaces $\partial/\partial z$ For a non-conservative system equation (15), after setting $\delta \psi = \frac{\partial \psi_T}{\partial f_j}\delta f_j$, generalises to

$$\frac{d}{dz}\left(\frac{\partial L}{\partial\left(\frac{\partial f_i}{\partial z}\right)}\right) - \frac{\partial L}{\partial f_i} = 2\Re \int \left[R\frac{\partial \psi_T^*}{\partial f_j}dx\right] \qquad (16)$$

where L is the *real*, averaged, *conservative* Lagrangian and, as stated earlier, $\psi_T^*$ is a suitable trial function. This result is identical to other formulations [25-27] for non-conservative media and will be deployed below in attempt to reconcile such a Lagrangian approach with the exact simulations for the gyroelectric investigations in hand.

The choice of trial function is complicated by the fact that a number of distinctive classes of solutions is known to emerge in the non-gyroelectric case [15]. Nevertheless, it is useful to adopt the following

$$\psi_T(x,z) = \eta \operatorname{sech}(\rho(x-x_0))\exp\left(i\frac{\xi}{2}(x-x_0) - i\theta\right) \qquad (17)$$

in which the parameters $\eta, \rho, x_0, \xi, \theta$ are all functions of z and account, therefore, for many of the dynamical properties being sought.

The average Lagrangian is

$$L = \frac{\xi \eta^2}{\rho} \frac{dx_0}{dz} + \frac{2\eta^2}{\rho} \frac{d\theta}{dz} + \frac{1}{3}\rho\eta^2 + \frac{1}{4}\frac{\xi^2 \eta^2}{\rho}$$
$$- \frac{2}{3}\frac{\eta^4}{\rho} + \frac{16}{45}\frac{\upsilon\eta^6}{\rho} - \eta^2 \int_{-\infty}^{\infty} Q(x) \operatorname{sech}^2(\rho(x-x_0)) dx \qquad (18)$$

and the equations of motion are

$$\frac{d\xi}{dz} = -\rho \frac{d}{dx_0} \int Q(x) \operatorname{sech}^2(\rho(x-x_0)) dx - \frac{\xi}{2}\frac{\rho}{\eta^2} F_\theta + \frac{\rho}{\eta^2} F_{x_0} \qquad (19)$$

$$\frac{d\rho}{dz} = \frac{\rho^2}{\eta^2} \frac{F_\theta}{4} \qquad (20)$$

$$\frac{d\eta}{dz} = \frac{3}{4}\frac{\rho}{\eta} F_\theta \qquad (21)$$

$$\frac{dx_0}{dz} = -\frac{1}{2}\xi \qquad (22)$$

$$\frac{d\theta}{dz} = \frac{2}{3}\rho^2 + \frac{1}{3}\eta^2 - \frac{16}{15}\upsilon\eta^4 + \rho \int Q(x) \operatorname{sech}^2(\rho(x-x_0)) dx$$
$$- \frac{1}{4}\rho^2 \frac{d}{d\rho} \int Q(x) \operatorname{sech}^2(\rho(x-x_0)) dx \qquad (23)$$

in which

$$F_\theta \equiv 2\Re e\left(\int R \frac{\partial \psi^*}{\partial \theta} dx\right) = \frac{8}{3}\beta\rho\eta^2 + \beta\frac{\xi^2\eta^2}{\rho} - \frac{8}{3}\varepsilon\frac{\eta^4}{\rho} + \frac{32}{15}\mu\frac{\eta^6}{\rho} + 4\frac{\alpha\eta^2}{\rho} \qquad (24)$$

$$F_{x_0} = 2\Re e\left(\int R \frac{\partial \psi^*}{\partial x_0} dx\right) = 2\beta\xi\rho\eta^2 + \frac{1}{2}\beta\frac{\xi^3\eta^2}{\rho} - \frac{4}{3}\varepsilon\frac{\xi\eta^4}{\rho} + \frac{16}{15}\mu\frac{\xi\eta^6}{\rho} + 2\alpha\frac{\xi\eta^2}{\rho} \qquad (25)$$

These results will be used in the next section in an attempt to map the Lagrangian dynamics onto the exact simulations.

## IV.   Numerical results

**Voigt configuration**

This is a 1D case and it is sketched in Fig.1, where it is shown that magneto-optic control can be achieved through the application of an electric current I. The latter is carried by a simple wire that produces a magnetic field that quickly drops to zero in the $\pm x$ directions. This arrangement is designed to create an inhomogeneous magnetic field that saturates the magnetization under the wire and falls away from that value along each x-direction. Typically a material like $(LuNdBi)_3(FeAl)_5O_{12}$ on a $Gd_3Ga_5O_{12}$ substrate can be used [3] for the magneto-optic component and some form of gain/loss material exhibiting a cubic-quintic nonlinearity is used to support the spatial solitons. To proceed any further a form of $Q_1(x)$ is required to apply a magnetic field influence upon the spatial solitons. Fortunately, a hyperbolic tangent function model is very close to the observed magnetization versus magnetic field behaviour of magnet-optic materials. Since that is the case, $Q_1(x)$=Atanh(KH/$H_s$) is assumed here, where A and K are constants. K is selected to make $\tanh \to 1$ in the higher magnetic field regions, H is the applied magnetic field value and $H_s$ is the saturation magnetic field. A further detail is that a real wire has a finite thickness but that only serves to broaden the x-distribution of the magnetic field, so for simplicity the wire will be assumed to be infinitely thin in the examples reported here. Directly under the wire, the magnetic field is H=I/($2\pi$d), where d is the waveguide thickness, and at some distance down the x axis the field is $H=I/[2\pi\sqrt{d^2+x^2}]$. Typically, the magneto-optic material saturates at 300Oe = 23.87mA/m, so a convenient form for the magneto-optic function [3] is

$$Q_1(x) = A\tanh\left( K \frac{I(mA)}{2\pi\sqrt{x(\mu m)^2 + d(\mu m)^2}} \frac{1}{23.87} \right) \qquad (26)$$

where $Q_1(x)$ reverses sign as the current direction reverses. In Fig.1 the current flows along the negative z axis to provide a magnetic field pointing along the positive x-axis. Hence the current flow is in the opposite direction to the propagation direction, which is along +z. The applied electric current is measured in mA and x and d are measured in $\mu m$. In order to gauge the paradigm influence of the use of such an electrode structure, attached to the kind of waveguide sketched in Fig.1 a current of $\pm 200$mA is used and d is taken to be the order of $1\mu m$. In this case A=2 but this can be varied by adopting a new set of waveguide parameters. K=1 in the examples below but again this can be assigned to a higher or lower values to permit $\tanh \to 1$ in regions of high field.

(A) *periodic breathing*

There are several interesting cases that emerge from the work of Akhmediev and co-workers [13-17] so the data sets they adopted will be adhered to here so that reference points can be established before the magnetic field is introduced. One of the outcomes form the cubic-quintic complex Ginzburg-Landau model is the appearance of pulsating solitons that engage in periodic breathing [17] as they propagate. As pointed out in the literature [13], this dynamical system can be elegantly represented by constructing a 2D plot of $|\psi(0,z)|^2$, the intensity at x=0, versus the energy $E = \int |\psi(x,z)|^2 dx$ for all points along the z-axis. This treatment of the dynamical system produces a kind of limit cycle that will be displayed below.

The first simulation studies the behaviour of beams that are super Gaussian or sech-shaped at the input plane. The initial beams have the shapes:

$$\psi(x, z=0) = \psi_0 \exp\left(-\left(\frac{x}{15}\right)^8\right); \quad \psi_0 = \sqrt{2} \tag{27}$$

$$\psi(x, z=0) = \psi_0 \sech\left(\frac{x}{w}\right); \quad w=1; \psi_0 = 1 \tag{28}$$

As expected, the Gaussian beam exhibits some significant early transient behaviour, until the final breathing form is achieved. For an initial sech-type input the final state is reached more rapidly. Fig.2 shows these features in parts (a) and (d) that represents soliton development in the absence of an applied magnetic field. For each type of input "breathing" is in a periodic manner. The role of the magnetic field is introduced through the $Q_1(x)\psi$ term and Figs. 2(b),(e) are evidence of what happens when I=200mA > 0 and $Q_1(x) > 0$, in which case the self-focusing is deepened. For the sech input the periodicity is suppressed by the magnetic field. If the input excitation is a super-Gaussian there is an initial attempt by the beam to maintain some pulsation but eventually the magnetic field prevails, and a narrow self-focussed stationary state is developed. For I < 0, $Q_1(x) < 0$, the defocusing is exacerbated and in both the cases shown in Figs.2(c),(f) the beam spreads significantly over a relatively short propagation distance. These results show that this system is impressively non-reciprocal. Fig.3 shows the x=0 intensity versus energy plane. For zero applied field, and a sech input, there is a closed loop and the points will keep on repeating themselves i.e. they will always lie upon this loop. In this sense it can be called a cycle. Upon the application of the kind of applied magnetic field distribution defined earlier, a well-defined stationary state is created for I >0 and this is represented by a single point on Fig.3(a). For I < 0 the plot degenerates to a straight line which is implied by Fig. 2((c). For a super-Gaussian input, there is also a closed loop for I = 0 but for I > 0 a spiralling path is taken, until a limit point is reached that represents the stable solution. This is another way to illustrate the non-reciprocal nature of the system in the presence of the magnetic field.

(B) *soliton explosions*

If he parameters are changed yet again to the set recommended in the literature [17] an her fascinating outcome is revealed and displayed in Fig.4. For both sech and super-Gaussian inputs remarkable phenomena have been discussed [13] in which the solitons are formed but they then explode at regular intervals as they progress down the z-axis and yet they reconstitute themselves between the explosions. The effect of an applied magnetic field upon this type of behaviour is shown. For I >0 the behaviour is very much like the I = 0 case because the self-focussing is being enhanced. There is a slight effect of the applied field because the effective potential well has been altered, which causes the solitons to "wobble" slightly. For I < 0 the defocusing is enhanced but nevertheless for this set of parameters the explosive pattern is still maintained even though they are now less frequent as the excitation passes along the z-axis. Here there is some evidence of radiation due to the defocusing nature of the magnetic field influence.

(C) *chaos*

Figure 5 shows how chaotic behaviour can ensue for a certain set of parameters and confirms even more graphically how rich the behaviour of solitons is for the complex Ginzburg-Landau equation. This chaotic behaviour has been carefully investigated in the absence of magneto-optic effects [13-18]. If the applied field is zero clearly defined, chaotic effects occur and this would be expressed on the dynamical 2D figure as a dense set of tracks that "wander about" within a closed region. The first thing to notice about the magnetic field action is that when I > 0 the system breaks up into a truly chaotic pattern. Reversing the current direction permits the defocusing to take hold and the beam spreads out very dramatically. For this set of parameters the role of the magnetic field is to produce a highly non-reciprocal behaviour which should have device potential.

(D) *Lagrangian numerical analysis*

The variational method will now be used to determine if the behaviour of solitons in the complex Ginzburg-Landau cubic-quintic equation will yield useful results. Two cases are selected as examples to demonstrate how well the Lagrangian works and these focus upon *moving* solitons and *breathing* solitons. For the moving solitons the aim is to use the applied magnetic field to control solitons previously classified [15] as "moving", in the sense that they moving at a finite velocity with respect to their frame of reference, so that the parameter set adopted is $\beta = 0.5$, $\alpha = 0.5$, $\nu = 0.1, \mu = 0.8, \varepsilon = 1.868$. In the case presented here the beam is located at x=-5 and the current wire is place at x=10, the current is I=200mA > 0. For this situation the solitons should see itself as being trapped into the potential well created by the magnetic field. This exactly what happens and the beam moves into a path dictated by the position of the current wire. This is clearly seen in Fig 6 where it is shown that the Lagrangian dynamics reproduce the full simulation behaviour very well.

Fig. 6 (a) shows that the beam becomes trapped in the neighbourhood of the wire and has been prevented from becoming established in the region that is accessible when $Q_1 = 0$.

For I< 0 the soliton is unable to cross the potential region because the latter will act as a barrier. This is precisely shown in Fig 6 (b).

z      z

z

Even for the breathing soliton good representation of the simulations, though the Lagrangian dynamics is achieved. Fig. 7 shows at least semi-quantitative agreement with the full simulations. In this case the current wire is placed at x=0 with a current of I =200 mA and the initial position of the beam is x=0, the dissipative parameters are $\beta = 0.08, \ \alpha = 0.1, \ \nu = 0.1, \mu = 0.1, \varepsilon = 0.66$

(E) *Faraday configuration: Optical vortices*

Retaining equation (8) in its 2D form enables it to be used for the study of optical vortices. Since these will propagate in the bulk the basic equation represents a circularly polarised beam in the Faraday configuration for materials like semimagnetic semiconductors. The initial excitation is

$$\psi(x,y,z=0) = \psi_0 \left(x + iy\,\text{sgn}(m)\right)^{|m|} \exp\left(-\left(\frac{r}{r_0}\right)^2\right) \quad (29)$$

$$r = \sqrt{x^2 + y^2}; \quad r_0 = 3; \ m = 1$$

This is the classic form [27,28] that contains a phase singularity at the origin. Its physical appearance is that of a Gaussian beam with a "hole" in it and m is the topological charge. In this case, there is no waveguide and the magneto-optic parameter is $Q_1 = \frac{\omega^2}{c^2} D_0^2 n^2 Q$, in which Q as function of x is assumed and for this example is set equal to $Q_1 = \pm 1.4 \sec h\left(\frac{x}{4}\right)$. Fig.9 shows that the action of the magnetic field is to transform the vortex into a bright pair of solitons. Fig.8 shows the input vortex and its phase distribution that reveals how the phase jumps by $2\pi$ as a circuit is made around the centre. The parameter $Q_1$ is only a function of x so there is a peak in this function that lies along the y-axis. This axis marks out therefore a potential well that bright spatial solitons can populate. As seen in Fig.7, this is precisely what happens and the vortex breaks down into two bright spatial solitons of opposite phase and they populate the y-axis. For the reversed direction of magnetic field the y-axis becomes a potential hill and the vortex breaks down into two bright solitons that reside on either side of it.

V.  Conclusions

This paper addresses the issue of how an applied magnetic field influences dissipative solitons. Two well-known configurations are adopted, namely the Voigt and the Faraday system. To consider the Voigt case, an asymmetric waveguide is created in which there is a nonlinear layer attached to a magnetooptic substrate. It is discovered that dramatic non-reciprocity is achieved by providing an inhomogeneous magnetic field distribution of the kind that can be supplied through a carefully chosen electrode structure. Some well-known data sets for the excitation of classes of dissipative solitons, using a complex Ginzburg-Landau cubic-quintic model are adopted. This means that soliton solutions that come under the headings pulsating, exploding and chaotic are investigated. It is demonstrated that the role of the magnetic field is to create enhanced potential wells or barriers and that this feature rapidly encourages self-focusing, or self-defocusing. A variational method is established for this non-conservative system and it is shown that the dynamical equations are in excellent agreement with the full simulations. Finally, a modest investigation into vortex propagation is presented in which it is shown that the externally applied magnetic field distribution causes an elementary vortex to dissolve into a pair of bright solitons that position themselves with respect to the potential created.

## VI. Acknowledgments

The authors would like to thank Dr Kiril Marinov, for useful discussions about the Lagrangian approach to dissipative systems.

## Figures

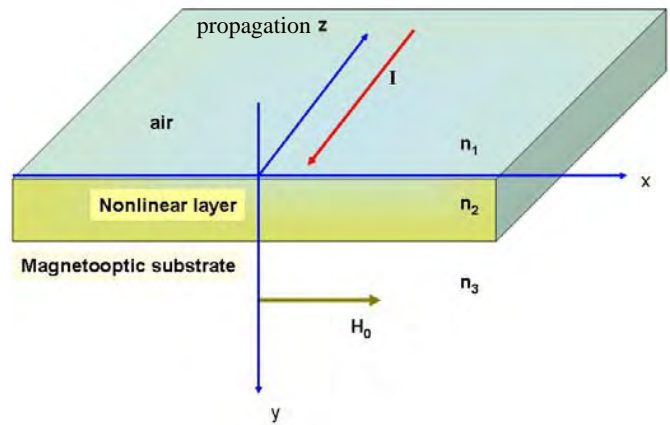

Fig 1. This shows the kind of asymmetric waveguide structure needed to ensure that the Voigt effect is O(Q), as opposed to O($Q^2$) in the bulk or for a symmetric waveguide. It consists of a nonlinear layer that will support spatial solitons and the substrate is a magneto-optic insulator with its magnetisation stimulated by a wire carrying a current I. The magnetic field component that creates the magnetisation is shown as $H_0$. The refractive indices are listed as $n_i$ and d is the thickness of the nonlinear layer.

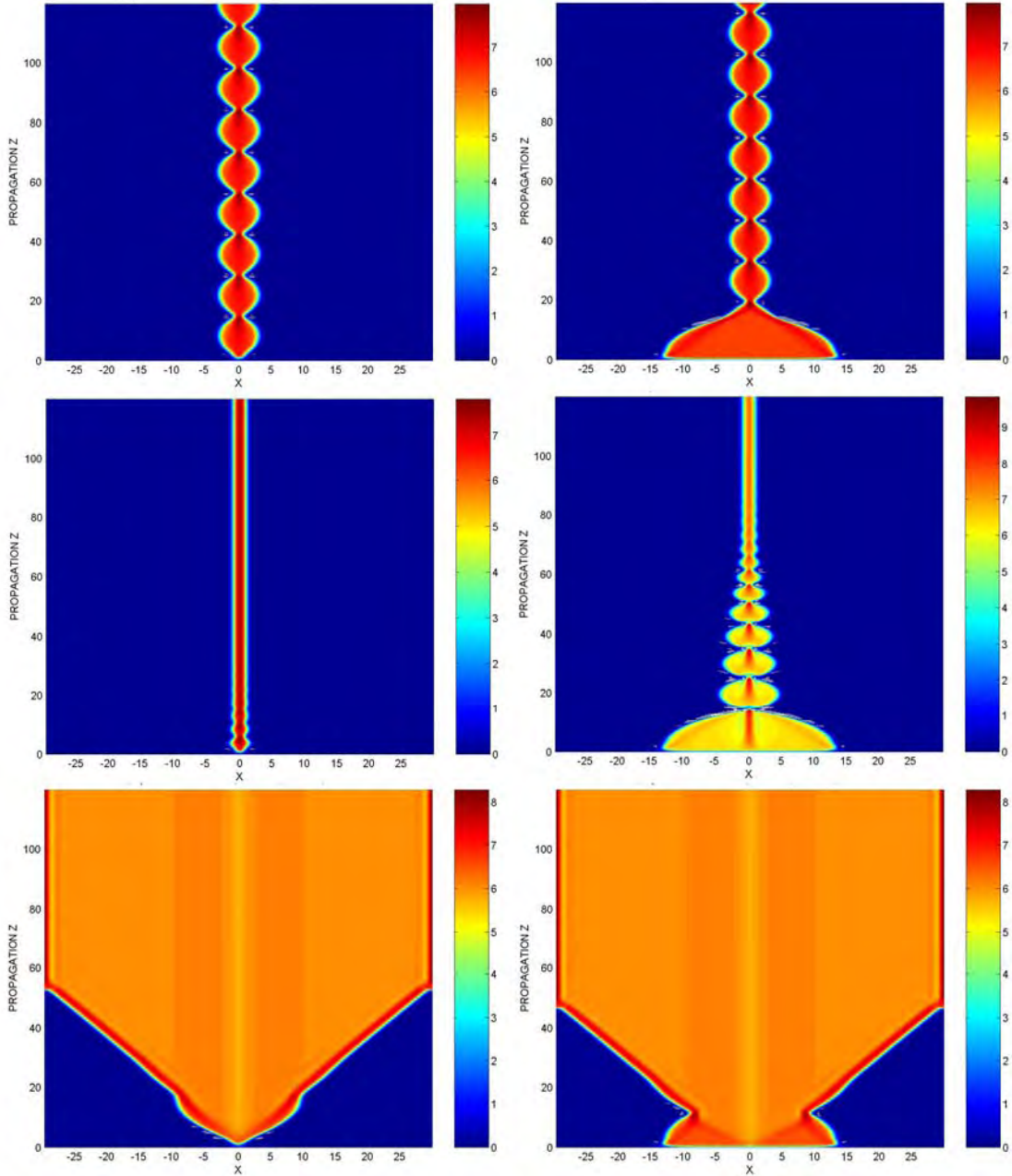

**Fig 2 Intensity plot of the beam propagation, for two different initial inputs : sech-type for the first column and super Gaussian type for the second column. The dissipative parameters are** $\beta = 0.08, \ \alpha = 0.1, \ \nu = 0.1, \mu = 0.1, \ \varepsilon = 0.66$. **The effect of the magnetic field is shown by setting the electric current as: I=0mA for the first row ( a and d); I =200mA for the second row (b and e ) and I=-200mA ( c and f) for the third row.**

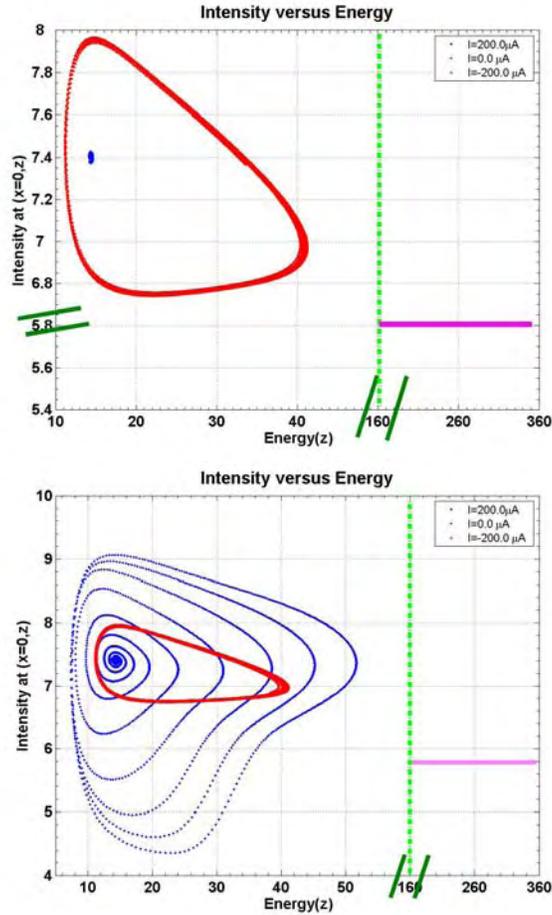

**Fig 3** Plots of the intensity versus energy when the initial beam is (a) sech-type (b) super Gaussian. The red curve corresponds to I=0mA, blue corresponds to I=200mA and the magenta curve to I=-200mA. The dissipative parameters are the same as those used in figure 2. Note that there are two different scales in each plot and the change points are designated by //.

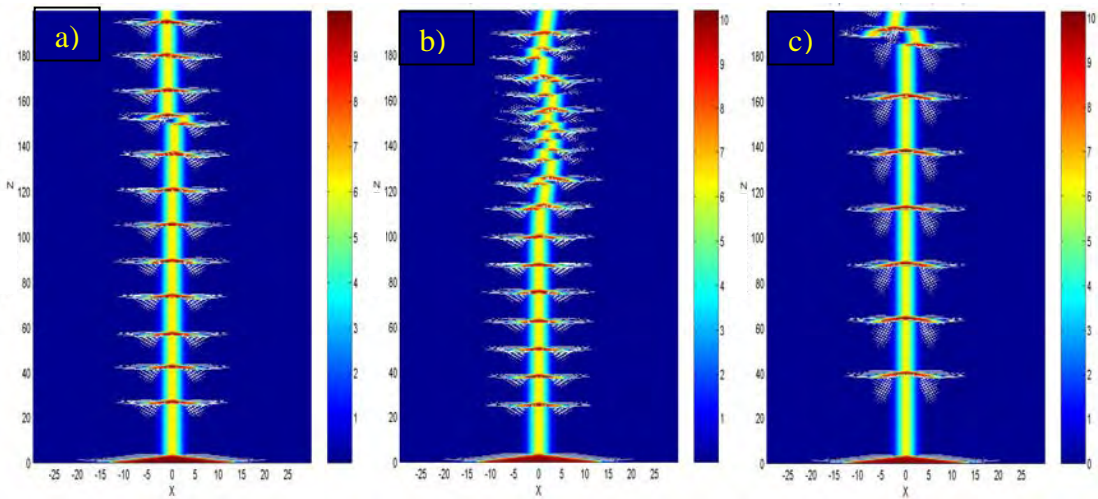

**Fig 4.** Intensity plot of an exploding soliton propagation under the influence of an external magnetic field. The dissipative parameters are $\beta = 0.125$, $\alpha = 0.1$, $\nu = 0.6, \mu = 0.1$, $\varepsilon = 1.0$. The effect of the magnetic field is displayed by setting the electric current as: **(a) I=0mA (b) I =200mA and (c) I=-200mA**

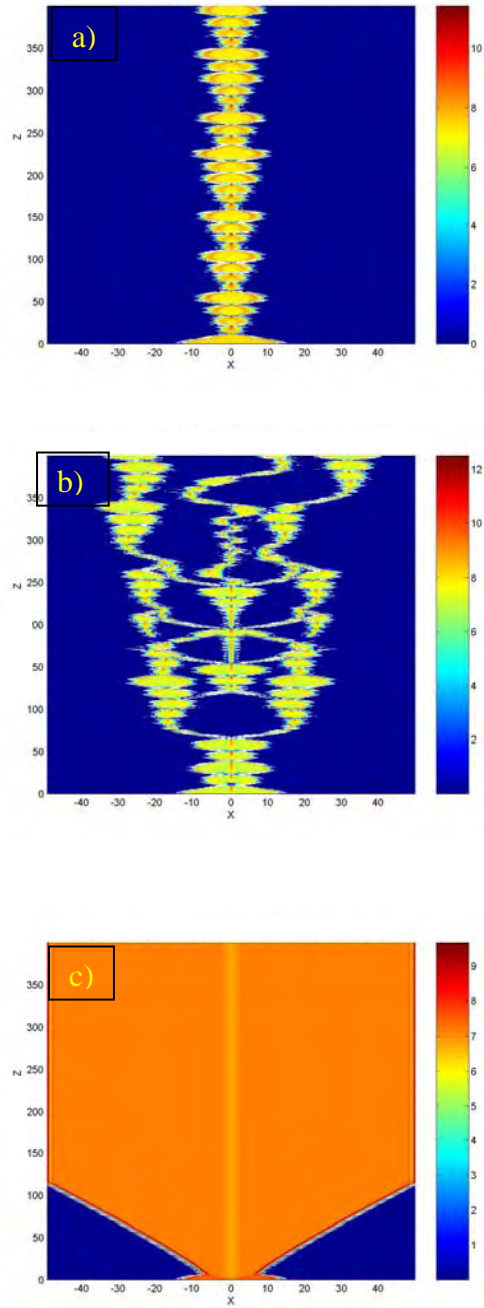

**Fig 5** Intensity plot of a "chaotic" soliton propagation under the influence of an external magnetic field. The dissipative parameters are $\beta = 0.04,\ \alpha = 0.1,\ \nu = 0.08, \mu = 0.1,\ \varepsilon = 0.75$. The effect of the magnetic field is shown by setting the electric current as: **(a) I=0mA (b) I =200mA and (c) I=-200mA**

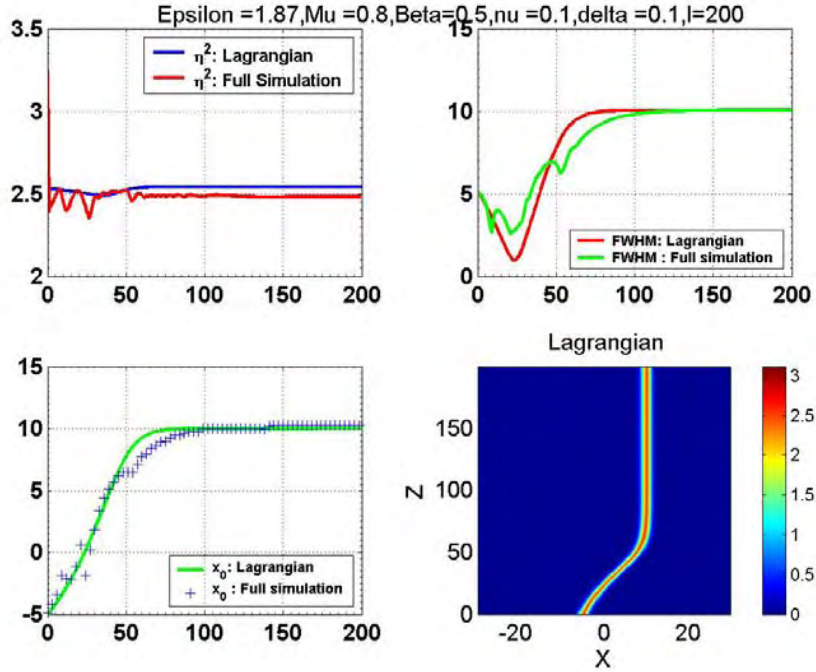

**Fig 6(a). Comparison of the Lagrangian results versus the full simulation, for the "moving soliton", initially centred at $x_0$=-5, under the influence of an external magnetic field provided by a electric current I=200mA located at $x_0$=10. The dissipative parameters are** $\beta = 0.5,\ \alpha = 0.5,\ \nu = 0.1, \mu = 0.8, \varepsilon = 1.868$. **The plots show the agreement between the Lagrangian and the data extracted from the full simulation. The figure shows the beam intensity (left upper corner), the FWHM (upper right corner) and the position of the beam (lower left corner). The bird's- eye view of the beam track (lower right corner) intensity colour plot of beam generated using the Lagrangian is also shown. Note that the beam is trapped in the region around x=10 and remains there because the magnetic field generates a deep well into which the beam 'falls'.**

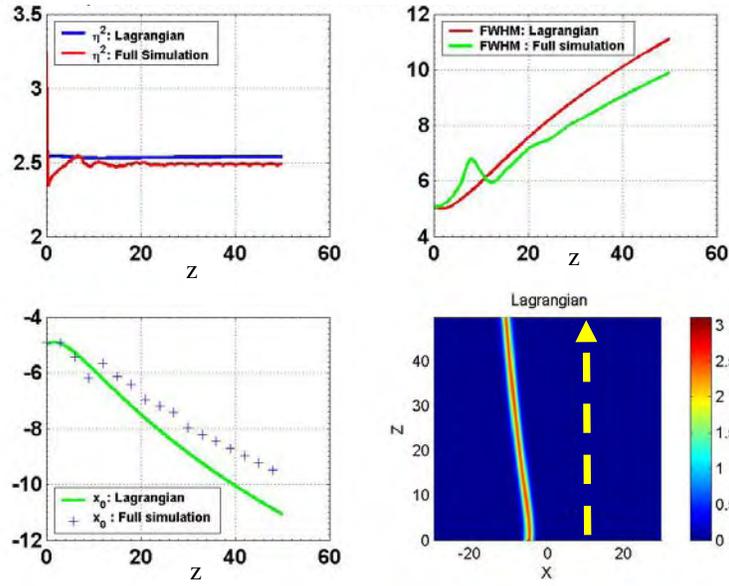

**Fig 6(b) Comparison of the Lagrangian results versus the full simulation, for the "moving soliton" initial centred at x$_0$=-5 under the influence of an external magnetic field provided by a electric current I=-200mA located at x$_0$=10. The dissipative parameters are $\beta = 0.5, \ \alpha = 0.5, \ \nu = 0.1, \mu = 0.8, \varepsilon = 1.868$, the plots show the agreement between the Lagrangian and the data extracted from the full simulation. The plot in the bottom right-hand corner shows that the soliton cannot access the right-hand side of the plane because the magnetic field generates a potential that behaves like a barrier.**

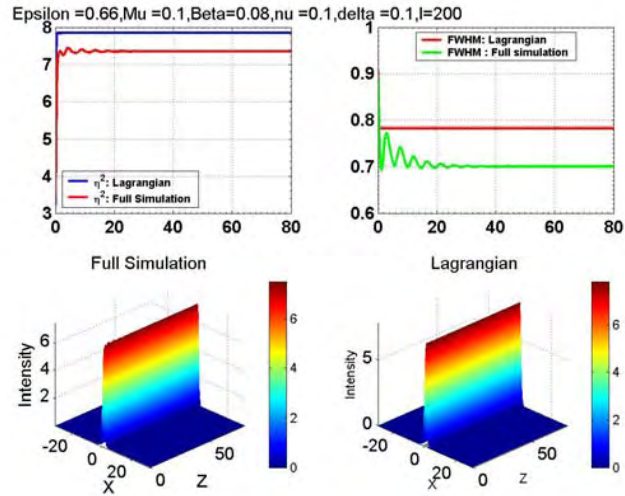

Figure 7 Comparison of the Lagrangian results versus the full simulation, for the "breathing soliton" initialy centred at x=-0 under the influence of an external magnetic field provided by a electric current I=200mA located at x=10, the dissipative parameters are $\beta = 0.08, \ \alpha = 0.1, \ \nu = 0.1, \mu = 0.1, \varepsilon = 0.66$, the plots show the agreement between the Lagrangian and the data extracted from the full simulation. For the beam intensity (left upper corner), the FWHM upper right corner, 3D intensity colour plot of beam, full simulation (lower left side) and generated using the parameters given by the Lagrangian (lower right corner).

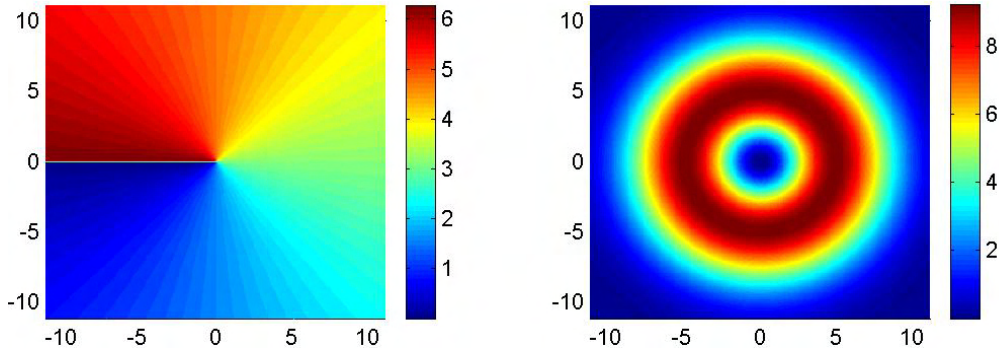

**Fig 8 a) Typical phase diagram for the beam at z=0. b) Intensity plot of an initial Gaussian beam, launched at z=0, with a singularity at the origin. The horizontal axis is x and the vertical is y.**

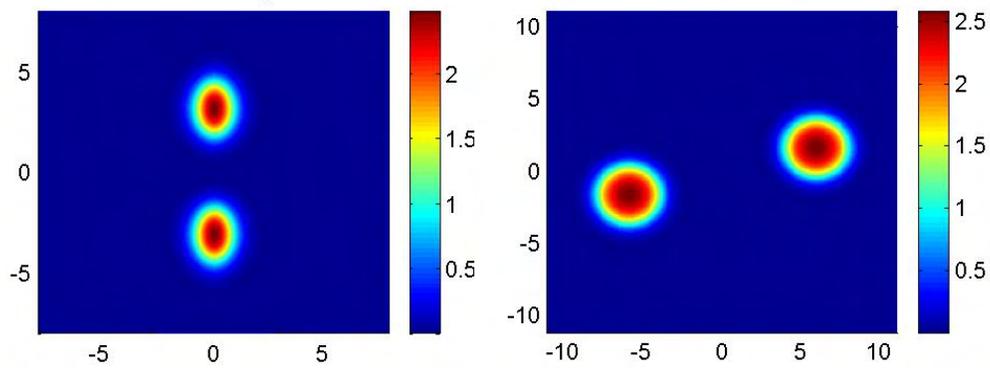

**Fig 9 Intensity plot of the beam propagation at z=60, when the magnetic field is switched on. (a) $Q_1>0$, (b) $Q_1 > 0$. The outcome in both cases is a pair of bright spatial solitons. The horizontal axis is x and the vertical is y.**